\documentclass[a4paper,10pt,twoside]{cpc-hepnp}
\usepackage{CJK,upgreek,fancyhdr}
\usepackage{multicol}
\usepackage{multirow}
\usepackage{graphicx}
\usepackage{booktabs}
\usepackage{amssymb,bm,mathrsfs,bbm,amscd}
\usepackage[tbtags]{amsmath}
\usepackage{lastpage}
\usepackage[english]{babel}
\usepackage{epsfig}
\usepackage{graphicx}
\usepackage{epstopdf}
\usepackage{longtable}

\begin{document}

\fancyhead[c]{\small Chinese Physics C~~~Vol. xx, No. x (2023) xxxxxx}
\fancyfoot[C]{\small 010201-\thepage}
\footnotetext[0]{Received \today}

\title{Inferring redshift and energy distributions of fast radio bursts from the first CHIME/FRB catalog\thanks{Supported by the National Natural Science Fund of China under grant Nos. 11873001, 12147102 and 12275034.}}

\author{%
Li Tang$^{1}$
\quad Hai-Nan Lin$^{2,3;1)}$\email{linhn@cqu.edu.cn}
\quad Xin Li$^{2,3}$
}
\maketitle

\address{%
$^1$ Department of Math and Physics, Mianyang Teachers' College, Mianyang 621000, China\\
$^2$ Department of Physics, Chongqing University, Chongqing 401331, China\\
$^3$ Chongqing Key Laboratory for Strongly Coupled Physics, Chongqing University, Chongqing 401331, China\\
}

\begin{abstract}
  We reconstruct the extragalactic dispersion measure \--- redshift relation (${\rm DM_E}-z$ relation) from well-localized fast radio bursts (FRBs) using Bayesian inference method. Then the ${\rm DM_E}-z$ relation is used to infer the redshift and energy of the first CHIME/FRB catalog. We find that the distributions of extragalactic dispersion measure and inferred redshift of the non-repeating CHIME/FRBs follow cut-off power law, but with a significant excess at the low-redshift range. We apply a set of criteria to exclude events which are susceptible to selection effect, but find that the excess at low redshift still exists in the remaining FRBs (which we call Gold sample). The cumulative distributions of fluence and energy for both the full sample and the Gold sample do not follow the simple power law, but they can be well fitted by the bent power law. The underlying physical implications remain to be further investigated.
\end{abstract}



\footnotetext[0]{\hspace*{-3mm}\raisebox{0.3ex}{$\scriptstyle\copyright$}2019
Chinese Physical Society and the Institute of High Energy Physics
of the Chinese Academy of Sciences and the Institute
of Modern Physics of the Chinese Academy of Sciences and IOP Publishing Ltd}%

\begin{multicols}{2}

\section{Introduction}\label{sec:introduction}

Fast radio bursts (FRBs) are energetic radio pulses of milliseconds duration happening in the Universe, see e.g. \cite{Petroff:2019tty,Cordes:2019cmq,Zhang:2020qgp,Xiao:2021omr} for recent review. The discovery of the first FRB can date back to 2007, when Lorimer et al. \cite{Lorimer:2007qn} reanalyzed the 2001 archive data of the Parkes 64-m telescope and found an anomalous radio pulse, which is now named FRB010724. Later on, Thornton et al. \cite{Thornton:2013iua} discovered several other similar radio pulses, which made FRBs receive great attention within the astronomy community. The origin of FRBs was still a mystery at that time, but the large dispersion measure (DM) implies that they are unlikely to originate from the Milky Way. The identification of host galaxy and the direct measurement of redshift confirmed that they have an extragalactic origin \cite{Keane:2016yyk,Chatterjee:2017dqg,Tendulkar:2017vuq}. Up to now, several hundreds of FRBs have been discovered \cite{Petroff:2016tcr,CHIMEFRB:2021srp}, among which only one is confirmed to originate from our Galaxy \cite{Andersen:2020hvz}. Phenomenological, FRBs can be divided into two types: repeaters and non-repeaters, according to whether they are one-off events or not. The majority of FRBs are apparently non-repeating, but it is still unclear if they will be repeating in the future. Most repeating FRBs are not very active, which repeat only two to three times \cite{Andersen:2019yex}. However, more than one thousand bursts have been observed from two extremely active sources, i.e. FRB20121102A \cite{Li:2021hpl} and FRB20201124A \cite{Xu:2021qdn}.

The physical origin of FRBs is still under extensive debate. Several theoretical models have been proposed to explain repeating and non-repeating FRBs, respectively. For example, giant pulses from young rapidly rotating pulsars \cite{Lyutikov:2016ueh}, the black hole battery model \cite{Mingarelli:2015bpo}, the ``Cosmic Comb" model \cite{Zhang:2017zse}, the inspiral and merger of binary neutron stars \cite{Totani:2013lia,Wang:2016dgs}, neutron star-white dwarf binary model \cite{Gu:2016ygt}, collision between neutron stars and asteroids \cite{Huang:2015peq}, highly magnetized pulsars travelling through asteroid belts \cite{Dai:2016qem,Liu:2020ynu}, young magnetars with fracturing crusts \cite{Suvorov:2019rzz}, axion stars moving through pulsar magnetospheres \cite{Buckley:2020fmh}, and so on. Although there is no standard model yet, it is widely accepted that the progenitor of FRB should at least involve one neutron star or magnetar. The recently discovered magnetar-associated burst in our Milky Way strongly supports the magnetar origin of some, if not all FRBs \cite{Andersen:2020hvz,Bochenek:2020zxn}. The statistical similarity between repeating FRBs and soft gamma repeaters further implies that they may have similar origin \cite{Wei:2021kdw,Sang:2021cjq}.

FRBs are energetic enough to be detectable up to high redshift, therefore they can be used as probes to investigate the cosmology \cite{Munoz:2016tmg,Yu:2017beg,Li:2017mek,Li:2020qei,Qiang:2019zrs,Wu:2020jmx,Wu:2021jyk,Lin:2021syj}, and to test the fundamental physics \cite{Wei:2015hwd,Wu:2016brq,Tingay:2016tgf,Bonetti:2016cpo,Wei:2021vvn}. Unfortunately, up to now most FRBs have no direct measurement of redshift. Although hundreds of FRBs have been measured, only a dozen of them are well localized. With such a small sample, we even do not clearly know the redshift distribution of FRBs. One way to solve this problem is to use the observed DM, which is an indicator of distance, to infer the redshift \cite{Zhang:2020ass,Hackstein:2020mxc,Hackstein:2021vkm,Qiang:2021ljr}. To this end, one should reasonably model the DM contribution from host galaxy and subtract it from the total observed DM. This is not an easy task, because too many factors may affect the host DM, such as the galaxy type, the inclination angle, the mass of host galaxy, the offset of FRB site from galaxy center, etc. A simple but rough assumption is that the host DM is a universal constant for all FRBs \cite{Yu:2017beg,Wu:2020jmx,Qiang:2021ljr}. Alternatively, Luo et al. \cite{Luo:2018tiy} assumed that the host DM follows the star-formation rate (SFR) of the host galaxy. However, Lin et al. \cite{Lin:2022afm} found no strong correlation between host DM and SFR from the limited sample of localized FRBs. A more reasonable way to deal with the host DM is to model it using a proper probability distribution and marginalize over the free parameters \cite{Macquart:2020lln,Zhang:2020mgq,Wu:2021jyk}. For example, Macquart et al. \cite{Macquart:2020lln} assumed that the host DM follows log-normal distribution, and reconstructed the DM-redshift relation from five well-localized FRBs. However, due to the small data sample, the DM-redshift relation has large uncertainty. As the discovery of more and more FRBs in recent years, it is interesting to recheck the DM-redshift relation and use it to infer the redshift of FRBs which has no direct measurement of spectroscopic or photometric redshift.

In this paper, we assume that the host DM of FRBs follows log-normal distribution, and reconstruct the DM-redshift relation from well localized FRBs using Bayesian inference method. Then the DM-redshift relation is used to infer the redshift of the first CHIME/FRB catalog \cite{CHIMEFRB:2021srp}. We further use the inferred redshift to calculate the isotropic energy of the CHIME/FRBs. The rest parts of this paper are arranged as follows: In Section 2, we reconstruct the DM-redshift relation from well-localized FRBs. In Section 3, we investigate the redshift and energy distributions of CHIME/FRBs. Finally, discussion and conclusions are given in Section 4.

\section{The DM-redshift relation from localized FRBs}\label{sec:host_frbs}

The interaction of electromagnetic waves with plasma leads to the frequency-dependent light speed. This plasma effect, although small, may cause detectable time delay between electromagnetic waves of different frequencies, if it is accumulated at cosmological distance. This phenomenon is more obvious for low-frequency electromagnetic waves, such as the radio wave as is observed in e.g. FRBs. The time delay between low- and high-frequency electromagnetic waves propagating from a distant source to earth is proportional to the integral of electron number density along the line-of-sight, i.e. the so called dispersion measure (DM). The observed DM of an extragalactic FRB can generally be decomposed into four main parts: the Milky Way interstellar medium (${\rm DM_{MW}}$), the Galactic halo (${\rm DM_{halo}}$), the intergalactic medium (${\rm DM_{IGM}}$), and the host galaxy (${\rm DM_{host}}$) \cite{Deng:2013aga,Gao:2014iva,Macquart:2020lln},
\begin{equation}\label{eq:DM_obs}
  {\rm DM_{obs}}={\rm DM_{MW}}+{\rm DM_{halo}}+{\rm DM_{IGM}}+\frac{{\rm DM_{host}}}{1+z},
\end{equation}
where ${\rm DM_{host}}$ is the DM of host galaxy in the FRB source frame, and the factor $1+z$ arises from the cosmic expansion. Occasionally, the ${\rm DM_{halo}}$ term is ignored, but this term is comparable to, or even larger than the ${\rm DM_{MW}}$ term for FRBs at high Galactic latitude.

The Milky Way ISM term (${\rm DM_{MW}}$) can be well modeled from pulsar observations, such as the NE2001 model \cite{Cordes:2002wz} and the YMW16 model \cite{Yao_2017msh}. For FRBs at high Galactic latitude, both models give consistent results. However, it is pointed out that the YMW16 model may overestimate ${\rm DM_{MW}}$ at low Galactic latitude \cite{KochOcker:2021fia}. Therefore, we use the NE2001 model to estimate the ${\rm DM_{MW}}$ term. The Galactic halo term (${\rm DM_{halo}}$) is not well constrained yet, and Prochaska $\&$ Zheng \cite{Prochaska:2019mkd} estimated that it is about in the range $50\sim 80~{\rm pc~cm^{-3}}$. Here we follow Macquart et al. \cite{Macquart:2020lln} and assume a conservative estimation on it, i.e. ${\rm DM_{halo}}=50~{\rm pc~cm^{-3}}$. The concrete value of ${\rm DM_{halo}}$ should not strongly affect our results, as its uncertainty is much smaller than the uncertainties of the ${\rm DM_{IGM}}$ and ${\rm DM_{host}}$ terms described bellow. Therefore, the first two terms on the right-hand-side of equation (\ref{eq:DM_obs}) can be subtracted from the observed ${\rm DM_{obs}}$. For convenience, we define the extragalactic DM as
\begin{equation}\label{eq:DM_E}
  {\rm DM_E}\equiv {\rm DM_{obs}}-{\rm DM_{MW}}-{\rm DM_{halo}}={\rm DM_{IGM}}+\frac{{\rm DM_{host}}}{1+z}.
\end{equation}

Given a specific cosmological model, the ${\rm DM_{IGM}}$ term can be calculated theoretically. Assuming that both hydrogen and helium are fully ionized \cite{Meiksin:2007rz,Becker:2010cu}, the ${\rm DM_{IGM}}$ term can be written in the standard $\Lambda$CDM model as \cite{Deng:2013aga,Zhang:2020ass}
\begin{equation}\label{eq:DM_IGM}
  \langle{\rm DM_{IGM}}(z)\rangle=\frac{21cH_0\Omega_bf_{\rm IGM}}{64\pi Gm_p}\int_0^z\frac{1+z}{\sqrt{\Omega_m(1+z)^3+\Omega_\Lambda}}dz,
\end{equation}
where $f_{\rm IGM}$ is the fraction of baryon mass in IGM, $m_p$ is the proton mass, $H_0$ is the Hubble constant, $G$ is the Newtonian gravitational constant, $\Omega_b$ is the normalized baryon matter density, $\Omega_m$ and $\Omega_\Lambda$ are the normalized densities of matter (including baryon matter and dark matter) and dark energy, respectively. In this paper, we work in the standard $\Lambda$CDM model with the Planck 2018 parameters, i.e. $H_0=67.4~{\rm km~s^{-1}~Mpc^{-1}}$, $\Omega_m=0.315$, $\Omega_\Lambda=0.685$ and $\Omega_{b}=0.0493$ \cite{Aghanim:2018eyx}. The fraction of baryon mass in IGM can be tightly constrained by directly observing the budget of baryons in different states \cite{Fukugita:1997bi}, or observing the radio dispersion on gamma-ray bursts \cite{Inoue:2003ga}. All the observations show that $f_{\rm IGM}$ is about 0.84. Using five well-localied FRBs, Li et al. \cite{Li:2020qei} also obtained the similar result. Therefore, we fix $f_{\rm IGM}=0.84$ to reduce the freedom. The uncertainty of these parameters should not significantly affect our results, since they are much smaller than the variation of ${\rm DM_{IGM}}$ described bellow.

Note that equation (\ref{eq:DM_IGM}) should be interpreted as the mean contribution from IGM. Due to the large-scale matter density fluctuation, the actual value would vary around the mean. Theoretical analysis and hydrodynamic simulations show that the probability distribution for ${\rm DM_{IGM}}$ has a flat tail at large values, which can be fitted with the following function \cite{Macquart:2020lln,Zhang:2020xoc}
\begin{equation}
  p_{\rm IGM}(\Delta)=A\Delta^{-\beta}\exp\left[-\frac{(\Delta^{-\alpha}-C_0)^2}{2\alpha^2\sigma_{\rm IGM}^2}\right], ~~~\Delta>0,
\end{equation}
where $\Delta\equiv{\rm DM_{IGM}}/\langle{\rm DM_{IGM}}\rangle$, $\sigma_{\rm IGM}$ is the effective standard deviation, $\alpha$ and $\beta$ are related to the inner density profile of gas in haloes, $A$ is a normalization constant, and $C_0$ is chosen such that the mean of this distribution is unity. Hydrodynamic simulations show that $\alpha=\beta=3$ provides the best match to the model \cite{Macquart:2020lln,Zhang:2020xoc}, thus we fix these two parameters. Simulations also show that the standard deviation $\sigma_{\rm IGM}$ approximately scales with redshift as $z^{-1/2}$ in the redshift range $z\lesssim 1$ \cite{McQuinn:2013tmc,Jaroszynski:2018vgh}. The redshift-dependence of $\sigma_{\rm IGM}$ is still unclear at $z>1$, so we just simply extrapolate this relation to high-redshift region. Therefore, we follow Macquart et al. \cite{Macquart:2020lln} and parameterize it as $\sigma_{\rm IGM}=Fz^{-1/2}$, where $F$ is a free parameter.

Due to the lack of detailed observation on the local environment of FRB source, the host term ${\rm DM_{host}}$ is poorly known. It may range from several tens to several hundreds ${\rm pc~cm^{-3}}$. For example, Xu et al. \cite{Xu:2021qdn} estimated that the ${\rm DM_{host}}$ of the repeating burst FRB20201124A is in the range $10< {\rm DM_{host}}< 310~{\rm pc~cm}^{-3}$; Niu et al. \cite{Niu:2021bnl} inferred ${\rm DM_{host}}\approx 900~{\rm pc~cm}^{-3}$ for FRB20190520B. Numerical simulations show that the probability of ${\rm DM_{host}}$ follows the log-normal distribution \cite{Macquart:2020lln,Zhang:2020mgq},
\begin{eqnarray}\nonumber\label{eq:P_host}
p_{\rm host}({\rm DM_{host}}|\mu,\sigma_{\rm host})&=&\frac{1}{\sqrt{2\pi}{\rm DM_{host}}\sigma_{\rm host}}\\
  &\times&\exp\left[-\frac{(\ln {\rm DM_{host}}-\mu)^2}{2\sigma_{\rm host}^2}\right],
\end{eqnarray}
where $\mu$ and $\sigma_{\rm host}$ are the mean and standard deviation of $\ln {\rm DM_{host}}$, respectively. This distribution has a median value of $e^\mu$ and variance $e^{\mu+\sigma_{\rm host}^2/2}(e^{\sigma_{\rm host}^2}-1)^{1/2}$. Theoretically, the log-normal distribution allows for the appearance of large value of ${\rm DM_{host}}$, as is shown by simulations, ${\rm DM_{host}}$ may be as large as $1000~{\rm pc~cm}^{-3}$ \cite{Hackstein:2020mxc}. Generally, the two parameters ($\mu,\sigma_{\rm host}$) may be redshift-dependent, but for non-repeating bursts they do not vary significantly with redshift \cite{Zhang:2020mgq}. For simplicity, we first follow Macquart et al. \cite{Macquart:2020lln} and treat them as two constant parameters. The possible redshift-dependence will be investigated later.

Given the distributions $p_{\rm IGM}$ and $p_{\rm host}$, the probability distribution of ${\rm DM_E}$ at redshift $z$ can be calculated as \cite{Macquart:2020lln}
\begin{eqnarray}\nonumber\label{eq:P_E}
  p_E({\rm DM_E}|z)&=&\int_0^{(1+z)\rm DM_E}p_{\rm host}({\rm DM_{host}}|\mu,\sigma_{\rm host})\\
  &\times&p_{\rm IGM}({\rm DM_E}-\frac{\rm DM_{host}}{1+z}|F,z)d{\rm DM_{host}}.
\end{eqnarray}
The likelihood that we observe a sample of FRBs with ${\rm DM_{E,\it i}}$ at redshift $z_i$ ($i=1,2,3,...,N$) is given by
\begin{equation}
  \mathcal{L}({\rm FRBs}|F,\mu,\sigma_{\rm host})=\prod_{i=1}^Np_{E}({\rm DM_{E,\it i}}|z_i),
\end{equation}
where $N$ is the total number of FRBs. Given the FRB data ($z_i,{\rm DM_{E,\it i}}$), the posterior probability distribution of the parameters ($F,\mu,\sigma_{\rm host}$) is given according to Bayes theorem by
\begin{equation}
  P(F,\mu,\sigma_{\rm host}|{\rm FRBs})\propto\mathcal{L}({\rm FRBs}|F,\mu,\sigma_{\rm host})P_0(F,\mu,\sigma_{\rm host}),
\end{equation}
where $P_0$ is the prior of the parameters.

Up to now, there are 19 well-localized extragalactic FRBs that have direct identification of host galaxy and well measured redshift\footnote{The FRB Host Database, http://frbhosts.org/}. Among them, we ignore FRB20200120E and FRB20190614D. The former is very close to our Galaxy (3.6 Mpc), and the peculiar velocity dominates over the Hubble flow, hence has a negative redshfit $z=-0.0001$ \cite{Bhardwaj:2021xaa,Kirsten:2021llv}. The latter has no direct measurement of spectroscopic redshift, but has a photometric redshift $z\approx 0.6$ \cite{Law:2020cnm}. All the rest 17 FRBs have well measured spectroscopic redshift. The main properties of the 17 FRBs are listed in Table \ref{tab:host}, which will be used in the following to reconstruct the ${\rm DM_E}$-redshift relation.

\end{multicols}
\begin{table}[htbp]
\centering
\caption{\small{The properties of the Host/FRB catalog. Column 1: FRB name; Columns 2 and 3: the right ascension and declination of FRB source on the sky; Column 4: the observed DM; Column 5: the DM of the Milky Way ISM calculated using the NE2001 model; Column 6: the extragalactic DM calculated by subtracting ${\rm DM_{\rm MW}}$ and ${\rm DM_{\rm halo}}$ from the observed ${\rm DM_{\rm obs}}$, assuming ${\rm DM_{\rm halo}}=50~{\rm pc~cm^{-3}}$ for the Milky Way halo; Column 7: the spectroscopic redshift; Column 8: repeating or non-repeating; Column 9: the references.}}\label{tab:host}
{\begin{tabular}{ccccccccl} 
\hline\hline 
FRBs & RA & Dec & ${\rm DM_{obs}}$ & ${\rm DM_{MW}}$ & ${\rm DM_E}$ & $z_{\rm sp}$ & repeat? & reference\\
& [ $^{\circ}$ ] & [ $^{\circ}$ ] & [${\rm pc~cm^{-3}}$] & [${\rm pc~cm^{-3}}$] & [${\rm pc~cm^{-3}}$] & & \\
\hline
20121102A & $82.99$ & $33.15$ &557.00 &157.60 &349.40 &0.1927 & Yes & Chatterjee et al. \cite{Chatterjee:2017dqg}\\
20180301A & $93.23$ & $4.67$ &536.00 &136.53 &349.47 &0.3305 & Yes & Bhandari et al. \cite{Bhandari:2021pvj}\\
20180916B & $29.50$ & $65.72$ &348.80 &168.73 &130.07 &0.0337 & Yes & Marcote et al. \cite{Marcote:2020ljw}\\
20180924B & $326.11$ & $-40.90$ &362.16 &41.45 &270.71 &0.3214 & No & Bannister et al. \cite{Bannister:2019iju}\\
20181030A & $158.60$ & $73.76$ &103.50 &40.16 &13.34 &0.0039 & Yes & Bhardwaj et al. \cite{Bhardwaj:2021hgc}\\
20181112A & $327.35$ & $-52.97$ &589.00 &41.98 &497.02 &0.4755 & No & Prochaska et al. \cite{2019Sci...366..231P}\\
20190102C & $322.42$ & $-79.48$ &364.55 &56.22 &258.33 &0.2913 & No & Macquart et al. \cite{Macquart:2020lln}\\
20190523A & $207.06$ & $72.47$ &760.80 &36.74 &674.06 &0.6600 & No & Ravi et al. \cite{Ravi:2019alc}\\
20190608B & $334.02$ & $-7.90$ &340.05 &37.81 &252.24 &0.1178 & No & Macquart et al. \cite{Macquart:2020lln}\\
20190611B & $320.74$ & $-79.40$ &332.63 &56.60 &226.03 &0.3778 & No & Macquart et al. \cite{Macquart:2020lln}\\
20190711A & $329.42$ & $-80.36$ &592.60 &55.37 &487.23 &0.5217 & Yes & Macquart et al. \cite{Macquart:2020lln}\\
20190714A & $183.98$ & $-13.02$ &504.13 &38.00 &416.13 &0.2365 & No & Heintz et al. \cite{2020ApJ...903..152H}\\
20191001A & $323.35$ & $-54.75$ &507.90 &44.22 &413.68 &0.2340 & No & Heintz et al. \cite{2020ApJ...903..152H}\\
20191228A & $344.43$ & $-29.59$ &297.50 &33.75 &213.75 &0.2432 & No &Bhandari et al. \cite{Bhandari:2021pvj}\\
20200430A & $229.71$ & $12.38$ &380.25 &27.35 &302.90 &0.1608 & No & Bhandari et al. \cite{Bhandari:2021pvj}\\
20200906A & $53.50$ & $-14.08$ &577.80 &36.19 &491.61 &0.3688 & No & Bhandari et al. \cite{Bhandari:2021pvj}\\
20201124A & $77.01$ & $26.06$ &413.52 &126.49 &237.03 &0.0979 & Yes & Fong et al. \cite{Fong:2021xxj}\\
\hline
\end{tabular}}
\end{table}
\begin{multicols}{2}

\begin{figure}[htbp]
 \centering
 \includegraphics[width=0.45\textwidth]{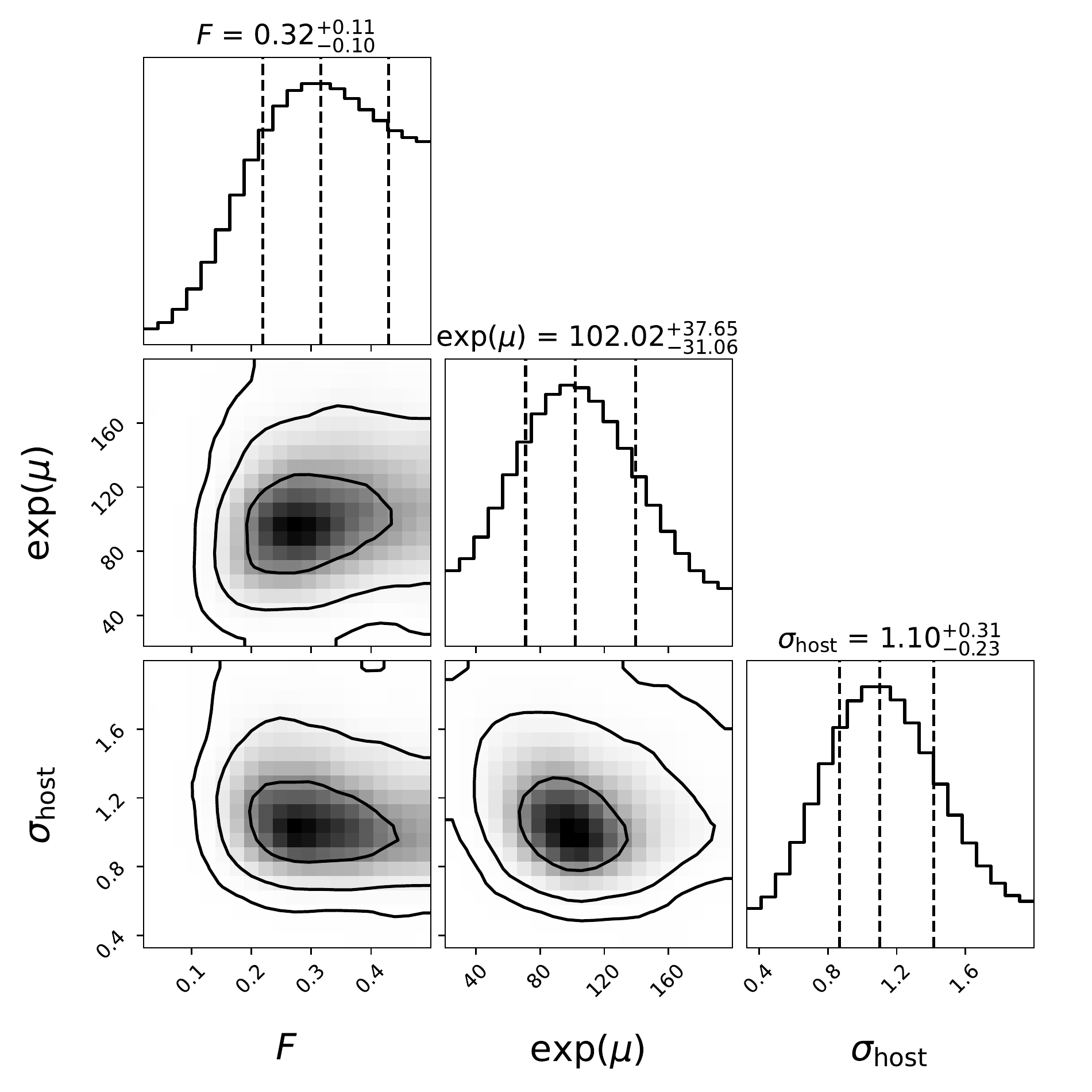}
 \includegraphics[width=0.45\textwidth]{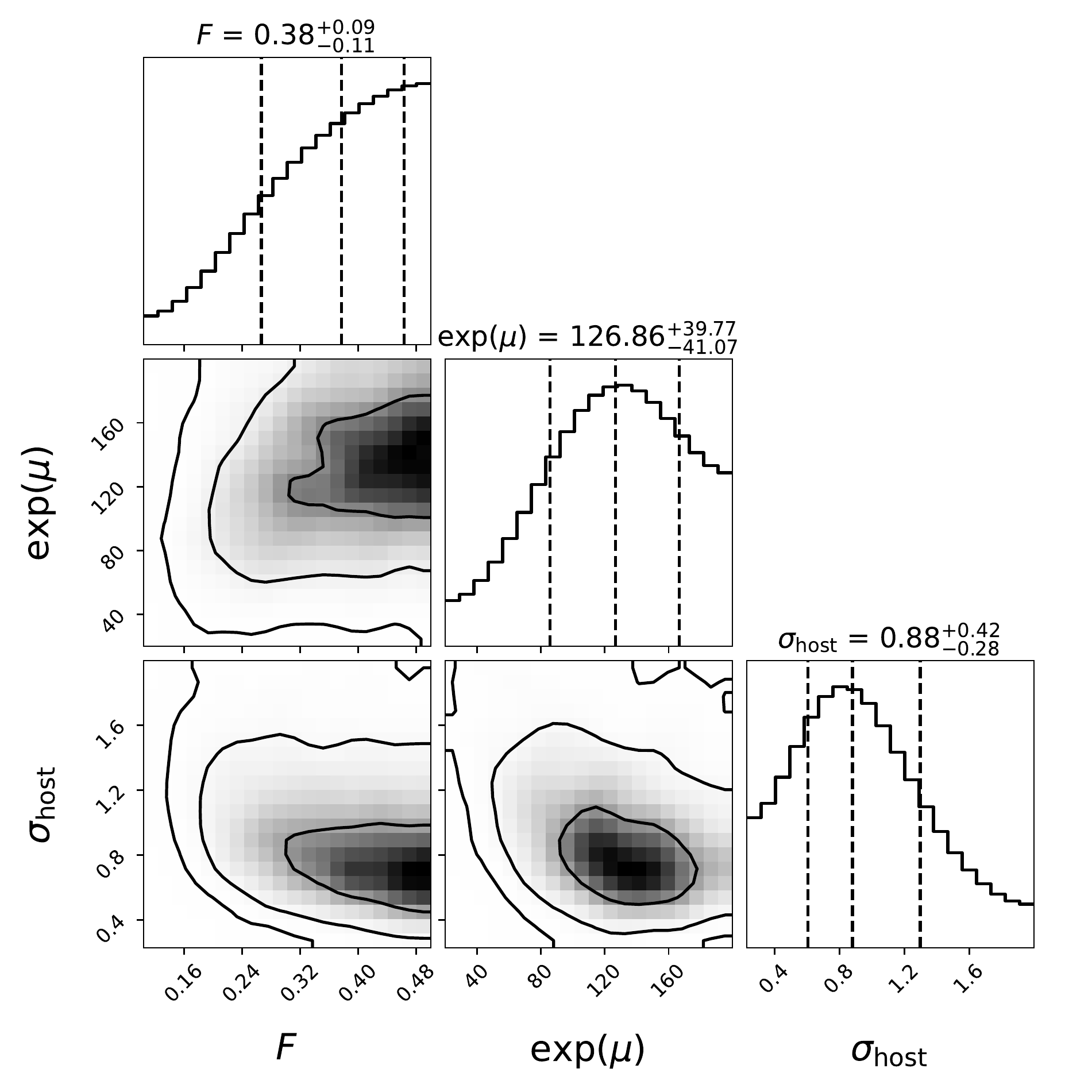}
 \caption{Constraints on the free parameters ($F$, $e^\mu, \sigma_{\rm host}$) using the full sample (left panel) and the non-repeaters (right panel). The contours from the inner to outer represent $1\sigma$, $2\sigma$ and $3\sigma$ confidence regions, respectively.}\label{fig:constrain}
\end{figure}

We first use the full 17 FRBs to constrain the free parameters ($F$,$e^\mu,\sigma_{\rm host}$). We use $e^\mu$ rather than $\mu$ as a free parameter, as was done in Macquart et al. \cite{Macquart:2020lln}, because the former directly represents the median value of ${\rm DM_{host}}$. The posterior probability density functions of the free parameters are calculated using the publicly available python package \textsf{emcee} \cite{Foreman-Mackey:2012any}, while the other cosmological parameters are fixed to the Planck 2018 values \cite{Aghanim:2018eyx}. The same flat priors as that in Macquart et al. \cite{Macquart:2020lln} are used for the free parameters: $F\in \mathcal{U}(0.01,0.5)$, $e^\mu\in \mathcal{U}(20,200)~{\rm pc~cm^{-3}}$ and $\sigma_{\rm host}\in \mathcal{U}(0.2,2.0)$. The posterior probability density functions and the confidence contours of the free parameters are plotted in the left panel of Figure \ref{fig:constrain}. The median values and $1\sigma$ uncertainties of the free parameters are $F=0.32_{-0.10}^{+0.11}$, $e^\mu=102.02_{-31.06}^{+37.65}~{\rm pc~cm^{-3}}$ and $\sigma_{\rm host}=1.10_{-0.23}^{+0.31}$.

\end{multicols}
\begin{multicols}{2}

\begin{figure}[htbp]
 \centering
 \includegraphics[width=0.48\textwidth]{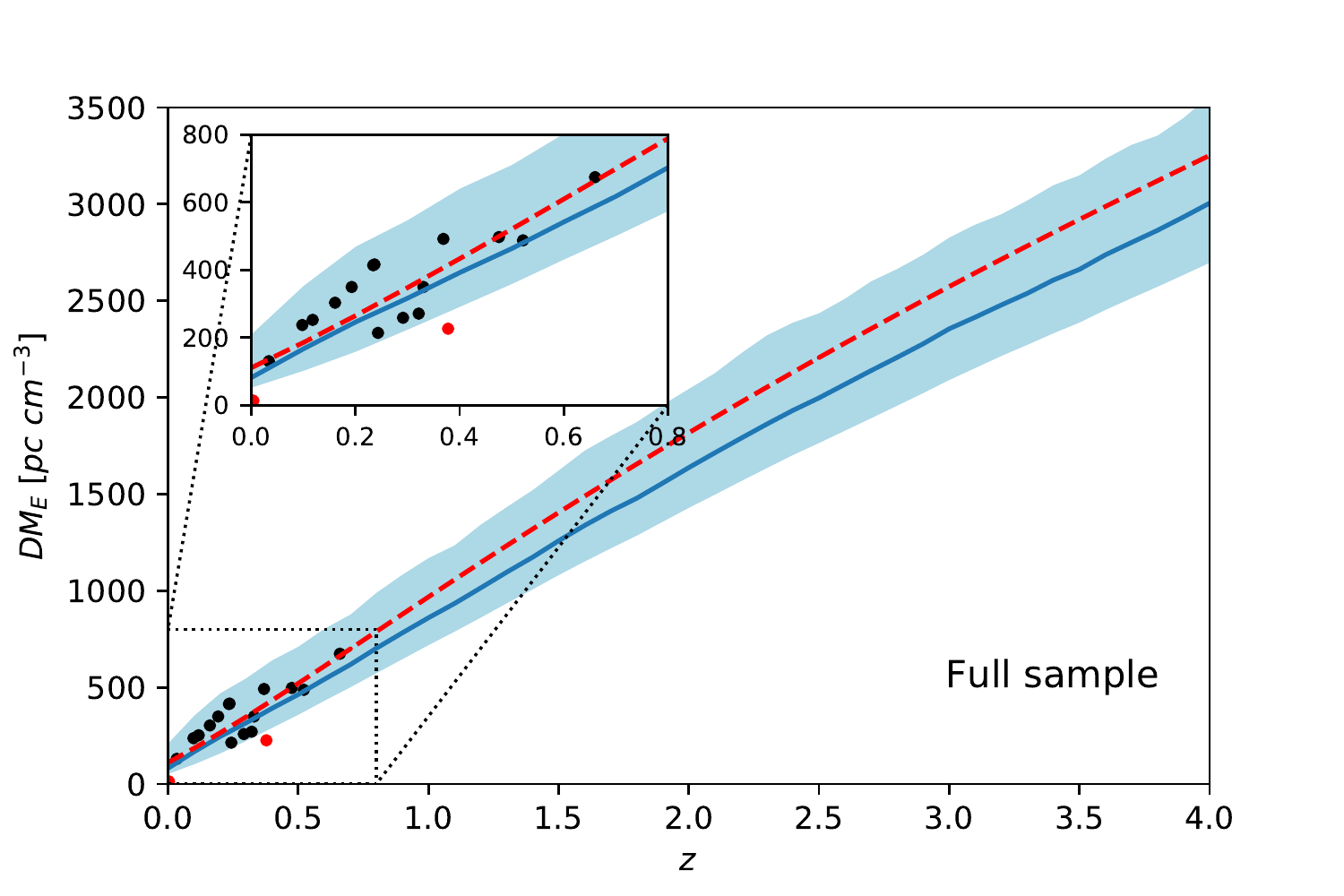}
 \includegraphics[width=0.48\textwidth]{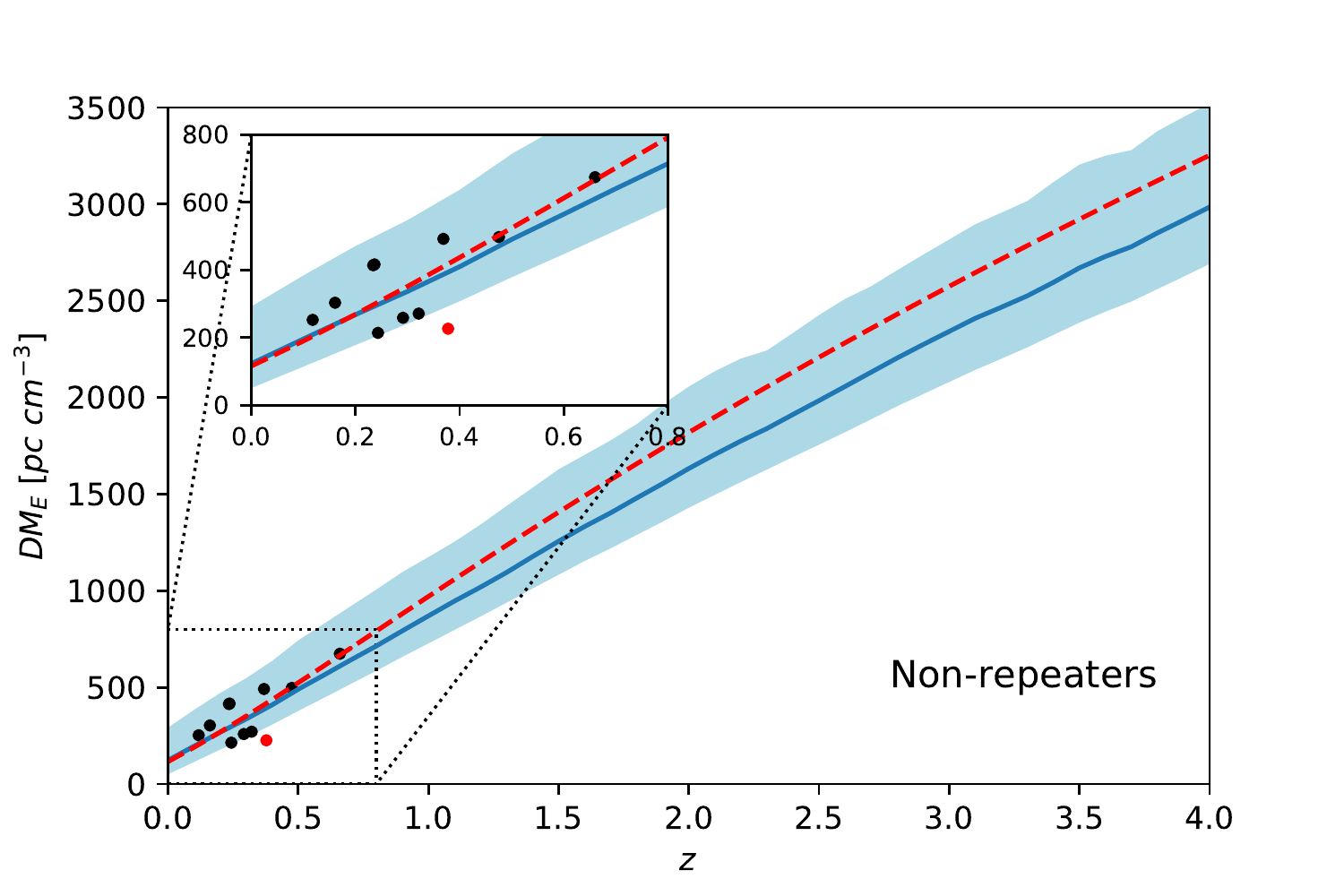}
 \caption{The ${\rm DM_E}-z$ relation obtained from full sample (left panel) and non-repeaters (right panel). The dark blue line is the median value and the light blue region is $1\sigma$ uncertainty. The dots are the FRB data points and the outliers are highlighted in red. The red-dashed line is the best-fitting result obtained using the least-$\chi^2$ method. The inset is the zoom-in of the low-redshift range.}\label{fig:DM_E}
\end{figure}

With the parameters ($F$, $e^\mu,\sigma_{\rm host}$) constrained, we calculate the probability distribution of ${\rm DM_E}$ at any redshift in the range $0<z<4$ according to equation (\ref{eq:P_E}). The reconstructed ${\rm DM_E}-z$ relation is plotted in the left panel of Figure \ref{fig:DM_E}. The dark blue line is the median value and the light blue region is the $1\sigma$ uncertainty. For comparison, we also plot the best-fitting curve by directly fitting equation (\ref{eq:DM_E}) to the FRB data using the least-$\chi^2$ method (the red-dashed line), where ${\rm DM_{IGM}}$ is replaced by its mean given in equation (\ref{eq:DM_IGM}). The least-$\chi^2$ method is equivalent to assume that both ${\rm DM_{IGM}}$ and ${\rm DM_{host}}$ follow Gaussian distribution around the mean. The least-$\chi^2$ curve gradually deviates from the median value of the reconstructed ${\rm DM_E}-z$ relation at high redshift, but due to the large uncertainty they are still consistent within $1\sigma$ uncertainty. We find that 15 out of the 17 FRBs well fall into the $1\sigma$ range of the reconstructed ${\rm DM_E}-z$ relation. Two outliers, FRB20181030A and FRB20190611B (the red dots in Figure \ref{fig:DM_E}), fall bellow the $1\sigma$ range of the ${\rm DM_E}-z$ relation, imply that the ${\rm DM_E}$ values of these two FRBs are smaller than expected. We note that the outlier FRB20181030A has a very smaller redshift $(z=0.0039)$ and a very low extragalactic DM (${\rm DM_E}=13.34~{\rm pc~cm^{-3}})$, so the peculiar velocity of its host galaxy couldn't be ignored. The redshift of the other outlier FRB20190611B is $z=0.3778$, and the observed DM of this burst is ${\rm DM_{obs}}=332.63~{\rm pc~cm^{-3}}$. The normal burst FRB20200906A has a redshift ($z=0.3688$) similar to FRB20190611B, but with a much larger DM (${\rm DM_{obs}}=577.8~{\rm pc~cm^{-3}}$). Note that both FRB20200906A and FRB20190611B are non-repeating, and the sky positions of these two bursts differ significantly. The large difference of ${\rm DM_{obs}}$ of these two bursts may be caused by e.g. the fluctuation of matter density in IGM, the variation of host DM, or different local environment of FRB source \cite{Niu:2021bnl,wang:2022ldk}.

\end{multicols}
\begin{multicols}{2}

The full FRB sample includes 11 non-repeating FRBs and 6 repeating FRBs. The repeaters and non-repeaters may have different ${\rm DM_{host}}$. To check this, we re-constrain the parameters ($F,e^\mu,\sigma_{\rm host}$) using the 11 non-repeating FRBs. The confidence contours and the posterior probability distributions of the parameter space are plotted in the right panel of Figure \ref{fig:constrain}. The median values and $1\sigma$ uncertainties of the free parameters are $F=0.38_{-0.11}^{+0.09}$, $e^\mu=126.86_{-41.07}^{+39.77}~{\rm pc~cm^{-3}}$ and $\sigma_{\rm host}=0.88_{-0.28}^{+0.42}$. We obtain a little larger $e^\mu$ value but a smaller $\sigma_{\rm host}$ value than that constrained from the full FRBs. But they are still consistent with $1\sigma$ uncertainty. The reconstructed ${\rm DM_E}-z$ relation using the non-repeating sample is shown in the right panel of Figure \ref{fig:DM_E}. FRB20190611B is still an outlier (the other outlier FRB20181030A is a repeater). The ${\rm DM_E}-z$ relations of the full sample and the non-repeaters are well consistent with each other, but the latter has a little larger uncertainty, especially at the low-redshift range.

\begin{figure}[htbp]
 \centering
 \includegraphics[width=0.48\textwidth]{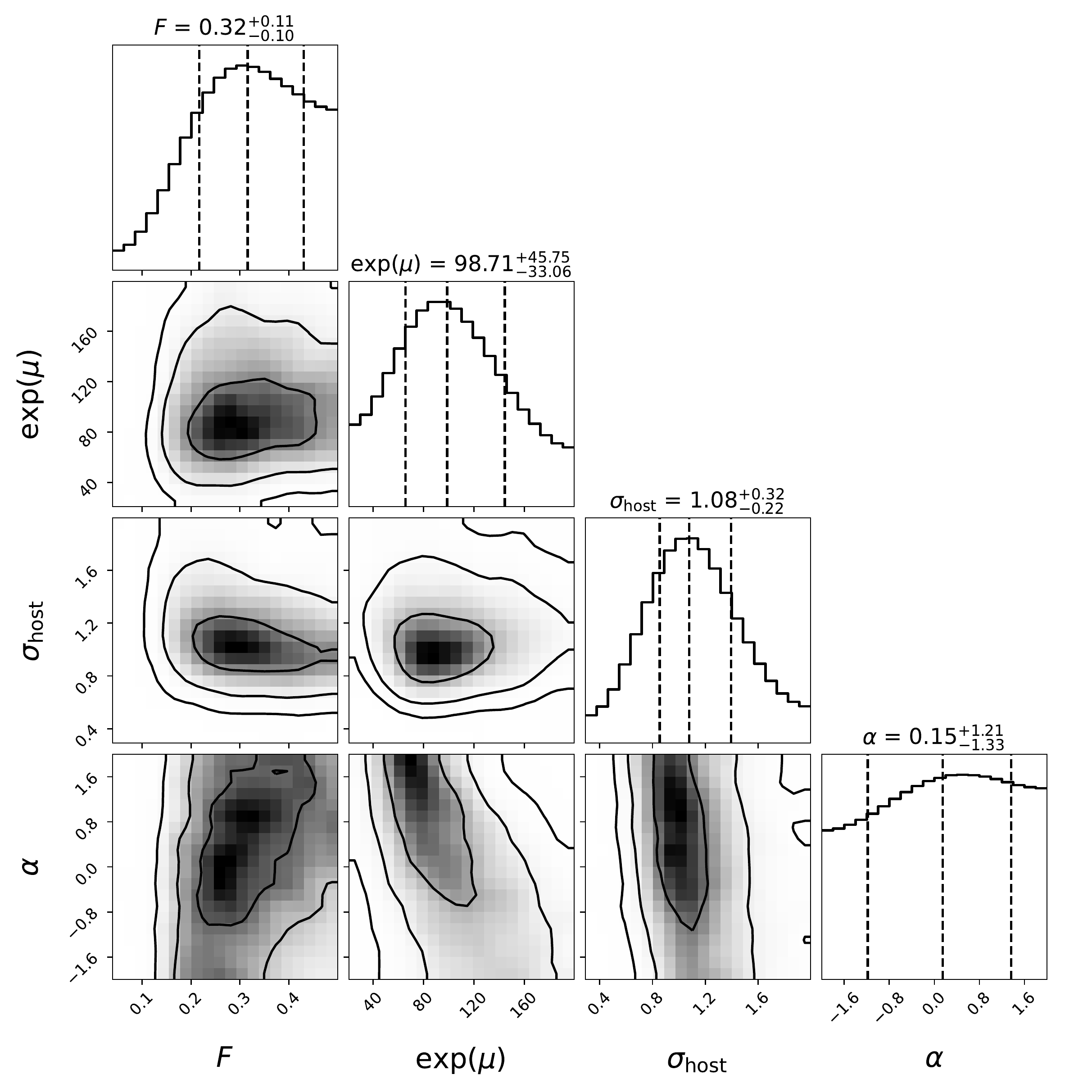}
 \includegraphics[width=0.48\textwidth]{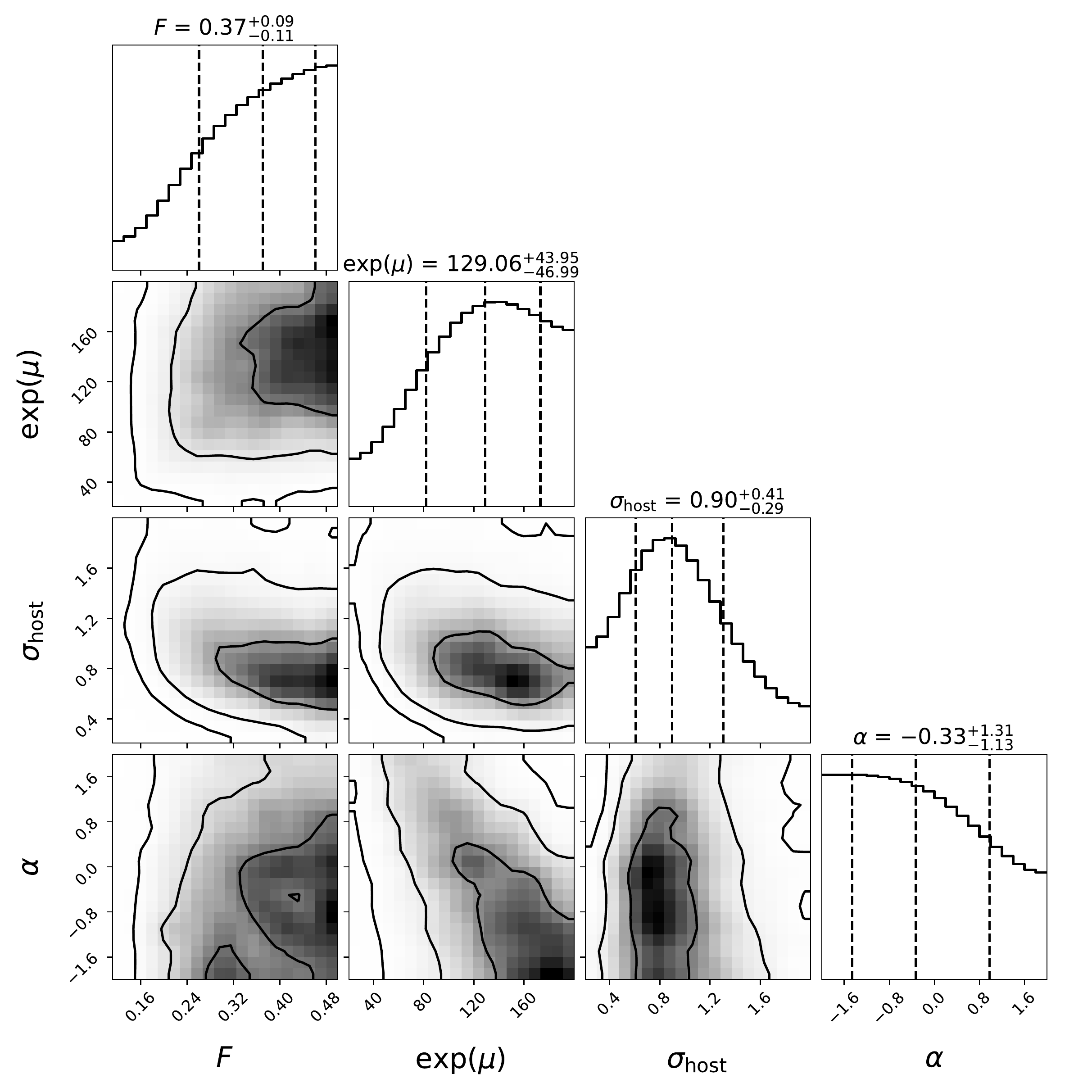}
 \caption{Constraints on the free parameters ($F, e^{\mu_0}, \sigma_{\rm host}, \alpha$) using the full sample (left panel) and the non-repeaters (right panel). The contours from the inner to outer represent $1\sigma$, $2\sigma$ and $3\sigma$ confidence regions, respectively.}\label{fig:constrain2}
\end{figure}

In general, $e^\mu$ and $\sigma_{\rm host}$ may evolve with redshift. Numerical simulations shows that the median value of ${\rm DM_{host}}$ has a power-law dependence on redshift, but $\sigma_{\rm host}$ does not change significantly \cite{Zhang:2020mgq}. To check this, we parameterize $e^{\mu}$ in the power-law form,
\begin{equation}
  e^\mu=e^{\mu_0}(1+z)^\alpha,
\end{equation}
and use the full FRB sample to constrain the parameters $(F, e^{\mu_0},\sigma_{\rm host},\alpha)$. Flat prior is adopted for $\alpha$ in the range $\alpha\in\mathcal{U}(-2,2)$. The posterior probability density functions and the confidence contours of the free parameters are plotted in the left panel of Figure \ref{fig:constrain2}. The best-fitting parameters are $F=0.32_{-0.10}^{+0.11}$, $e^{\mu_0}=98.71_{-33.06}^{+45.75}~{\rm pc~cm^{-3}}$, $\sigma_{\rm host} =1.08_{-0.22}^{+0.32}$ and $\alpha=0.15_{-1.33}^{+1.21}$. As can be seen, the parameter $\alpha$ couldn't be tightly constrained, while the constraints on the other three parameters are almost unchanged compared to the case when $\alpha=0$ fixed. This implies that there is no evidence for the redshift-dependence of $e^\mu$ with the present data. With the non-repeating FRBs, we arrive at the same conclusion (see the right panel of Figure \ref{fig:constrain2}). Therefore, it is safe to assume that $e^\mu$ is redshift-independent, at least in the low-redshift range $z<1$. But be caution that the universality of $e^\mu$ has not been proven at high redshift. Hence, the uncertainty on the ${\rm DM_E}-z$ relation in $z>1$ range may be underestimated.

\end{multicols}
\begin{multicols}{2}

\section{The redshift and energy distribution of CHIME/FRBs}\label{sec:chime_frbs}

\begin{figure}[htbp]
 \centering
 \includegraphics[width=0.48\textwidth]{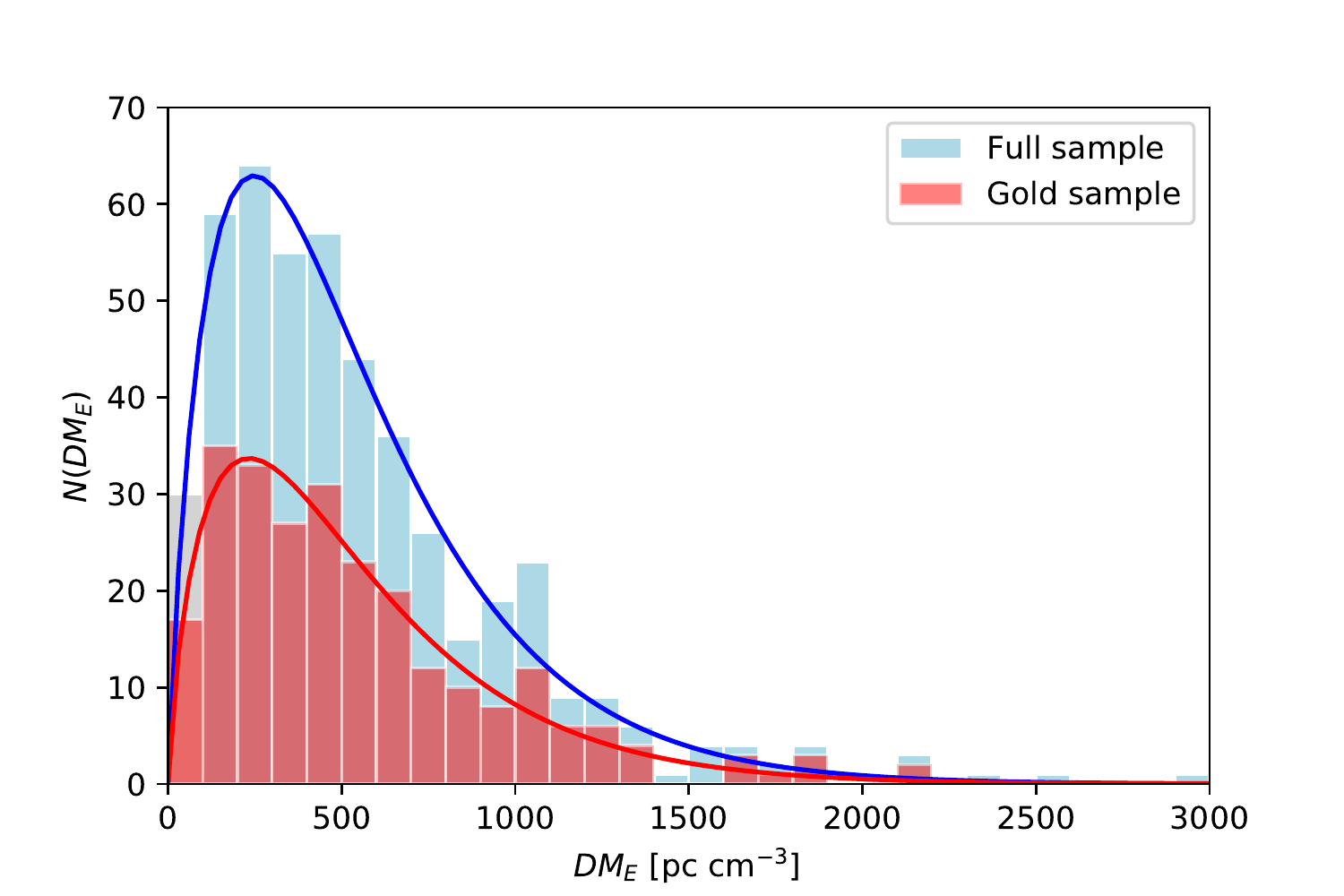}
 \includegraphics[width=0.48\textwidth]{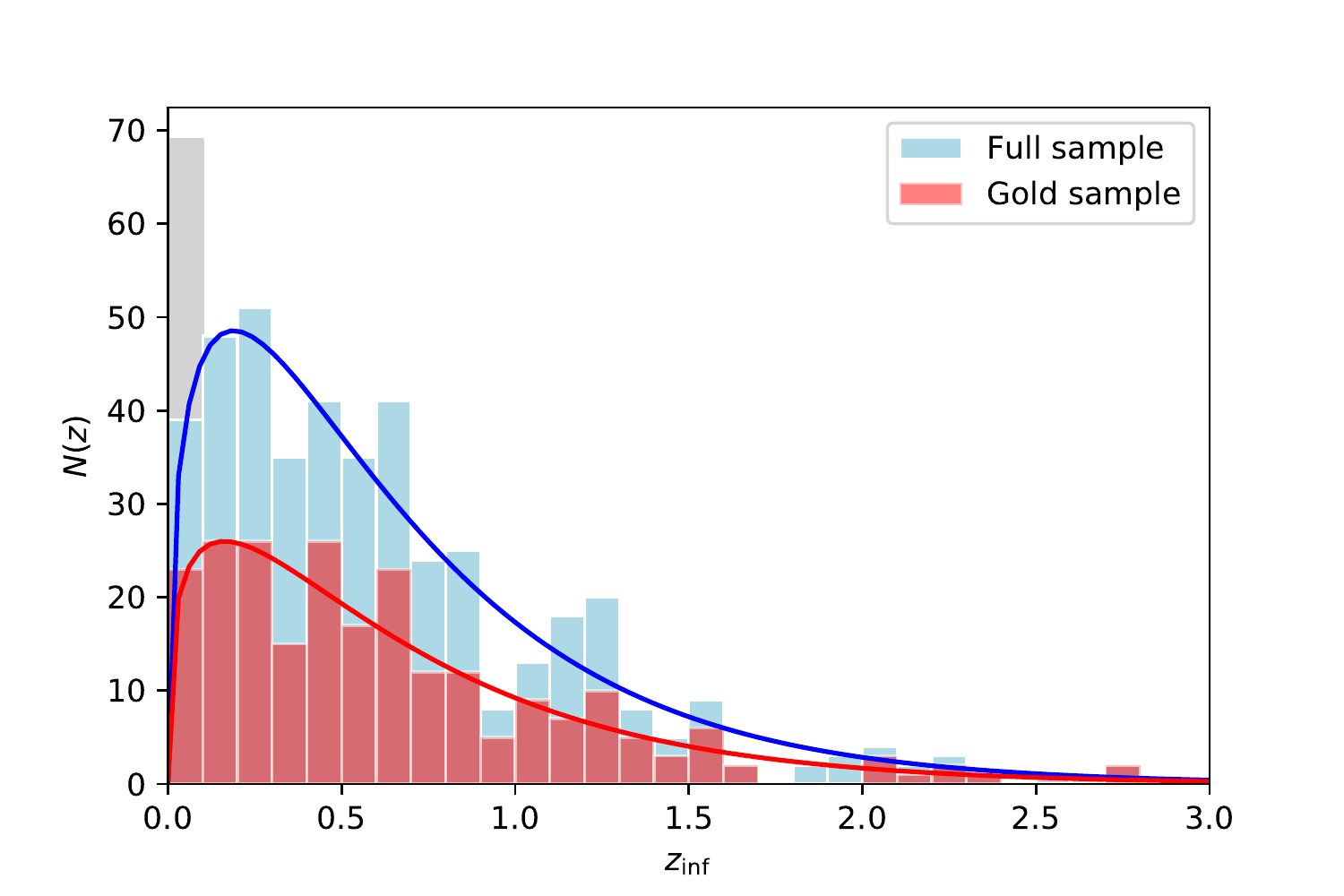}
 \caption{The histogram of ${\rm DM_E}$ (left panel) and inferred redshift (right panel) of the first non-repeating CHIME/FRB catalog. The left-most gray bar represent the 30 FRBs with ${\rm DM_E}<100~{\rm pc~cm^{-3}}$, which are expected to have $z<0.1$. The blue and red lines are the best-fitting CPL models for the Full sample and the Gold sample, respectively.}\label{fig:hist_chime}
\end{figure}

The first CHIME/FRB catalog consists of 536 bursts, including 474 apparently non-repeating bursts and 62 repeating bursts from 18 FRB sources \cite{CHIMEFRB:2021srp}. In this paper, we focus on the 474 apparently non-repeating bursts, whose properties are listed in a long table in the {\it online material}. All the bursts have well measured ${\rm DM_{obs}}$, but most of them have no direct measurement of redshift. We calculate the extragalactic ${\rm DM_E}$ by subtracting ${\rm DM_{MW}}$ and ${\rm DM_{halo}}$ from the observed ${\rm DM_{obs}}$, where ${\rm DM_{MW}}$ is calculated using the NE2001 model \cite{Cordes:2002wz}, and ${\rm DM_{halo}}$ is assumed to be $50~{\rm pc~cm^{-3}}$ \cite{Macquart:2020lln}. The ${\rm DM_E}$ values of the 474 apparently non-repeating bursts fall into the range $20\--3000$ pc cm$^{-3}$. Among them 444 bursts have ${\rm DM_E}>100$ pc cm$^{-3}$, while the rest 30 bursts have ${\rm DM_E}<100$ pc cm$^{-3}$. The mean and median values of ${\rm DM_E}$ are 557 and 456 pc cm$^{-3}$, respectively. We divide the ${\rm DM_E}$ of the full non-repeating bursts into 30 uniform bins, with bin width $\Delta{\rm DM_E}=100$ pc cm$^{-3}$, and plot the histogram in the left panel of Figure \ref{fig:hist_chime}. The distribution of ${\rm DM_E}$ can be well fitted by cut-off power law (CPL),
\begin{equation}\label{eq:CPL}
  {\rm CPL}:~~~N(x)\propto x^\alpha\exp\left(-\frac{x}{x_c}\right),~~~x>0,
\end{equation}
with the best-fitting parameters $\alpha=0.86\pm 0.07$ and $x_c=289.49\pm 17.90$ pc cm$^{-3}$. This distribution has a peak at $x_p=\alpha x_c\approx 250$ pc cm$^{-3}$, which is much smaller than the median value and mean value of ${\rm DM_E}$.

\end{multicols}
\begin{multicols}{2}

Now we use the ${\rm DM_E}-z$ relation reconstructed using the full sample (using the non-repeating sample does not significantly affect our results) to infer the redshift of the non-repeating CHIME/FRBs. For FRBs with ${\rm DM_E}<100$ pc cm$^{-3}$, the ${\rm DM_{host}}$ term may dominate over the ${\rm DM_{IGM}}$ term, hence a smaller uncertainty on ${\rm DM_{host}}$ may cause large bias on the estimation of redshift. Therefore, when inferring the redshift using ${\rm DM_E}-z$ relation, we only consider the FRBs with ${\rm DM_E}>100$ pc cm$^{-3}$. From the ${\rm DM_E}-z$ relation we can know that ${\rm DM_E}(z=0.1)=169.9_{-73.4}^{+196.9}$ pc cm$^{-3}$ ($1\sigma$ uncertainty). Therefore, FRBs with ${\rm DM_E}<100$ pc cm$^{-3}$ are expected to have redshift $z<0.1$, while the lower limit is unable to be determined. The inferred redshifts for FRBs with ${\rm DM_E}>100$ pc cm$^{-3}$ are given in the {\it online material}. The inferred redshifts span the range $z_{\rm inf}\in(0.023,3.935)$. Three bursts have inferred redshift larger than 3, i.e., FRB20180906B with $z_{\rm inf}=3.935_{-0.705}^{+0.463}$, FRB20181203C with $z_{\rm inf}=3.003_{-0.657}^{+0.443}$, and FRB20190430B with $z_{\rm inf}=3.278_{-0.650}^{+0.449}$.

We divide the redshift range $0<z<3$ into 30 uniform bins, with bin width $\Delta z=0.1$, and plot the histogram of the inferred redshift in the right panel of Figure \ref{fig:hist_chime}. The distribution of the inferred redshift can be fitted by CPL model given in equation (\ref{eq:CPL}). The best-fitting parameters are $\alpha=0.39\pm 0.09$ and $x_c=0.48\pm 0.06$. The distribution has a peak at $z_p=\alpha x_c\approx 0.19$. The mean and median value of this distribution are $0.67$ and $0.52$, respectively. Considering the FRBs with ${\rm DM_E}<100$ ${\rm pc~cm}^{-3}$ (30 FRBs in total), which are expected to have $z<0.1$, there is a large excess compared to the CPL model in the redshft range $z<0.1$ (see the left-most gray bar in Figure \ref{fig:hist_chime}). This may be caused by the selection effect, since the detector is more sensitive to the near FRBs.

Amiri et al. \cite{CHIMEFRB:2021srp} provided a set of criteria to exclude events which are unsuitable to be used in population analyses. (1) Events with $S/N < 12$ are excluded; (2) Events having ${\rm DM_{obs}} < 1.5{\rm max(DM_{NE2001}, DM_{YMW16})}$ are excluded; (3) Events detected in far sidelobes are excluded; (4) Events detected during non-nominal telescope operations are excluded; (5) Highly scattered events ($\tau_{\rm scat}>10$ ms) are excluded. We call the remaining FRBs the Gold sample, which consists of 253 non-repeating FRBs. We plot the distributions of ${\rm DM_E}$ and redshifts of the Gold sample (together with the Full sample) in Figure \ref{fig:hist_chime}. Similar to the Full sample, the distributions of ${\rm DM_E}$ and redshifts of the Gold sample can also be fitted by CPL model. The best-fitting CPL model parameters are summarized in Table \ref{tab:cpl_parameter}. We see that parameters are not significantly changed compared to the Full sample. Note that the redshift distribution of the Gold sample shown in the right panel of Figure 4 only contains the FRBs with ${\rm DM_E>100~pc~cm^{-3}}$ (236 FRBs). The Gold sample still contains 17 FRBs with ${\rm DM_E<100~pc~cm^{-3}}$, whose redshifts are expected to be $z<0.1$. So the low-redshift excess still exists in the Gold sample.

\end{multicols}
\begin{table}[htbp]
\centering
\caption{\small{The best-fitting CPL model parameters for the distributions of ${\rm DM_E}$ and redshift.}}\label{tab:cpl_parameter}
{\begin{tabular}{lll} 
\hline\hline 
${\rm DM_E}$ (Full) & $\alpha=0.86\pm 0.07$ & $x_c=289.49\pm 17.90~{\rm pc~cm^{-3}}$\\
\hline
${\rm DM_E}$ (Gold) & $\alpha=0.77\pm 0.09$ & $x_c=302.82\pm 23.92~{\rm pc~cm^{-3}}$\\
\hline
redshift (Full) & $\alpha=0.39\pm 0.09$ & $x_c=0.48\pm 0.06$\\
\hline
redshift (Gold) & $\alpha=0.31\pm 0.11$ & $x_c=0.52\pm 0.08$\\
\hline
\end{tabular}}
\end{table}
\begin{multicols}{2}

Given the redshift, the isotropic energy of a burst can be calculated as \cite{James:2021jbo}
\begin{equation}\label{eq:energy}
  E=\frac{4\pi d_L^2F\Delta\nu}{(1+z)^{2+\alpha}},
\end{equation}
where $d_L$ is the luminosity distance, $F$ is the average fluence, $\alpha$ is the spectral index ($F_\nu\propto \nu^\alpha$), and $\Delta\nu$ is the waveband in which the fluence is observed. The fluence listed in the first CHIME/FRB catalog is averaged over the $400-800$ MHz waveband, hence $\Delta\nu=400$ MHz. The spectra indices of some bursts are not clear. Macquart et al. \cite{Macquart:2018rsa} showed that for a sample of ASKAP/FRBs, $\alpha=-1.5$ provides a reasonable fit. Hence, we fix $\alpha=-1.5$ for all the bursts. Note that the fluence given in the CHIME/FRB catalog is lower limit, since the fluence is measured assuming each FRB is detected at the location of maximum sensitivity. So the energy calculated using equation (\ref{eq:energy}) is lower limit. With the inferred redshift, we calculate the isotropic energy in the standard $\Lambda$CDM cosmology with the Planck 2018 parameters \cite{Aghanim:2018eyx}. The uncertainty of energy propagates from the uncertainties of fluence and redshift. The results are given in the {\it online material}. The isotropic energy spans about five orders of magnitude, from $10^{37}$ erg to $10^{42}$ erg, with the median value $\sim 10^{40}$ erg. Three bursts have energy above $10^{42}$ erg, see Table \ref{tab:energetic_bursts}. The isotropic energy of the furthest burst, FRB20180906B, is about $4\times 10^{41}$ erg.

\end{multicols}
\begin{table}[htbp]
\centering
\caption{\small{The most energetic bursts with $E>10^{42}$ erg. Column 1: FRB name; Columns 2 and 3: the right ascension and declination of FRB source on the sky; Column 4: the observed DM; Column 5: the DM of the Milky Way ISM calculated using the NE2001 model; Column 6: the extragalactic DM calculated by subtracting ${\rm DM_{\rm MW}}$ and ${\rm DM_{\rm halo}}$ from the observed ${\rm DM_{\rm obs}}$, assuming ${\rm DM_{\rm halo}}=50~{\rm pc~cm^{-3}}$ for the Milky Way halo; Column 7: the observed fluence; Column 8: the inferred redshift; Column 9: the isotropic energy; Column 10: the flag for Gold sample (flag=1) or not (flag=0). Note that the uncertainty of energy may be underestimated due to the lack of well-localized FRBs at $z>1$.}}\label{tab:energetic_bursts}
{\begin{tabular}{cccccccccc} 
\hline\hline 
FRBs & RA & Dec & ${\rm DM_{obs}}$ & ${\rm DM_{MW}}$ & ${\rm DM_E}$ & Fluence & $z_{\rm inf}$ & $\log(E/{\rm erg})$ & flag \\
& [$^{\circ}$ ] & [$^{\circ}$ ] & [${\rm pc/cm^{3}}$] & [${\rm pc/cm^{3}}$] & [${\rm pc/cm^{3}}$] & [Jy ms] & & & \\
\hline
20181219B & $180.79$ & $71.55$ & $1950.7$ & $35.8$ & $1864.9$ & $27.00\pm22.00$ & $2.300_{-0.511}^{+0.357}$ & $42.405_{-0.962}^{+0.388}$ & $1$\\
20190228B & $50.01$ & $81.94$ & $1125.8$ & $81.9$ & $993.9$ & $66.00\pm32.00$ & $1.175_{-0.355}^{+0.205}$ & $42.170_{-0.633}^{+0.324}$ & $0$\\
20190319A & $113.43$ & $5.72$ & $2041.3$ & $109.0$ & $1882.3$ & $19.40\pm4.20$ & $2.325_{-0.516}^{+0.359}$ & $42.271_{-0.335}^{+0.214}$ & $1$\\
\hline
\end{tabular}}
\end{table}
\begin{multicols}{2}

\begin{figure}[htbp]
 \centering
 \includegraphics[width=0.48\textwidth]{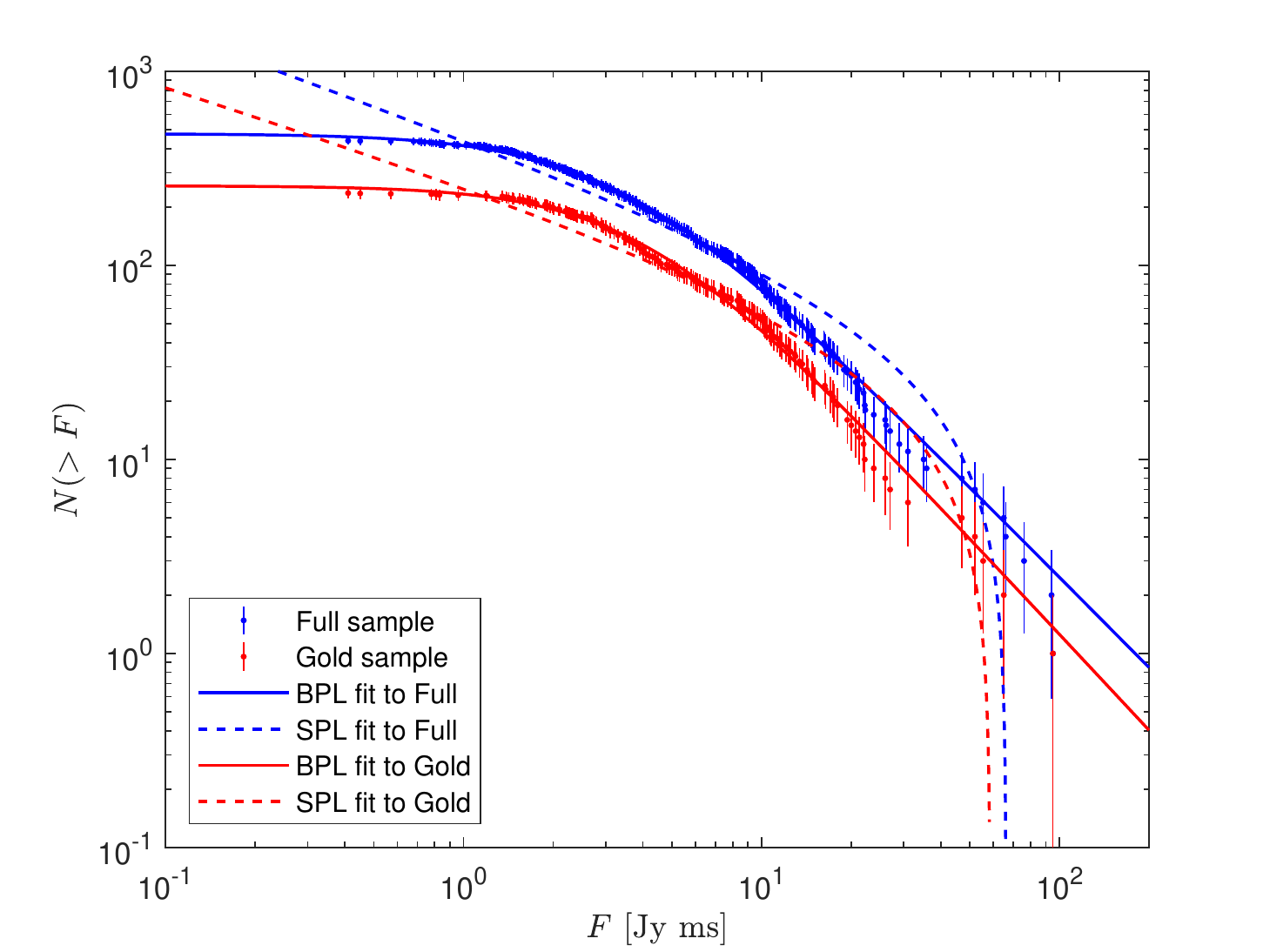}
 \includegraphics[width=0.48\textwidth]{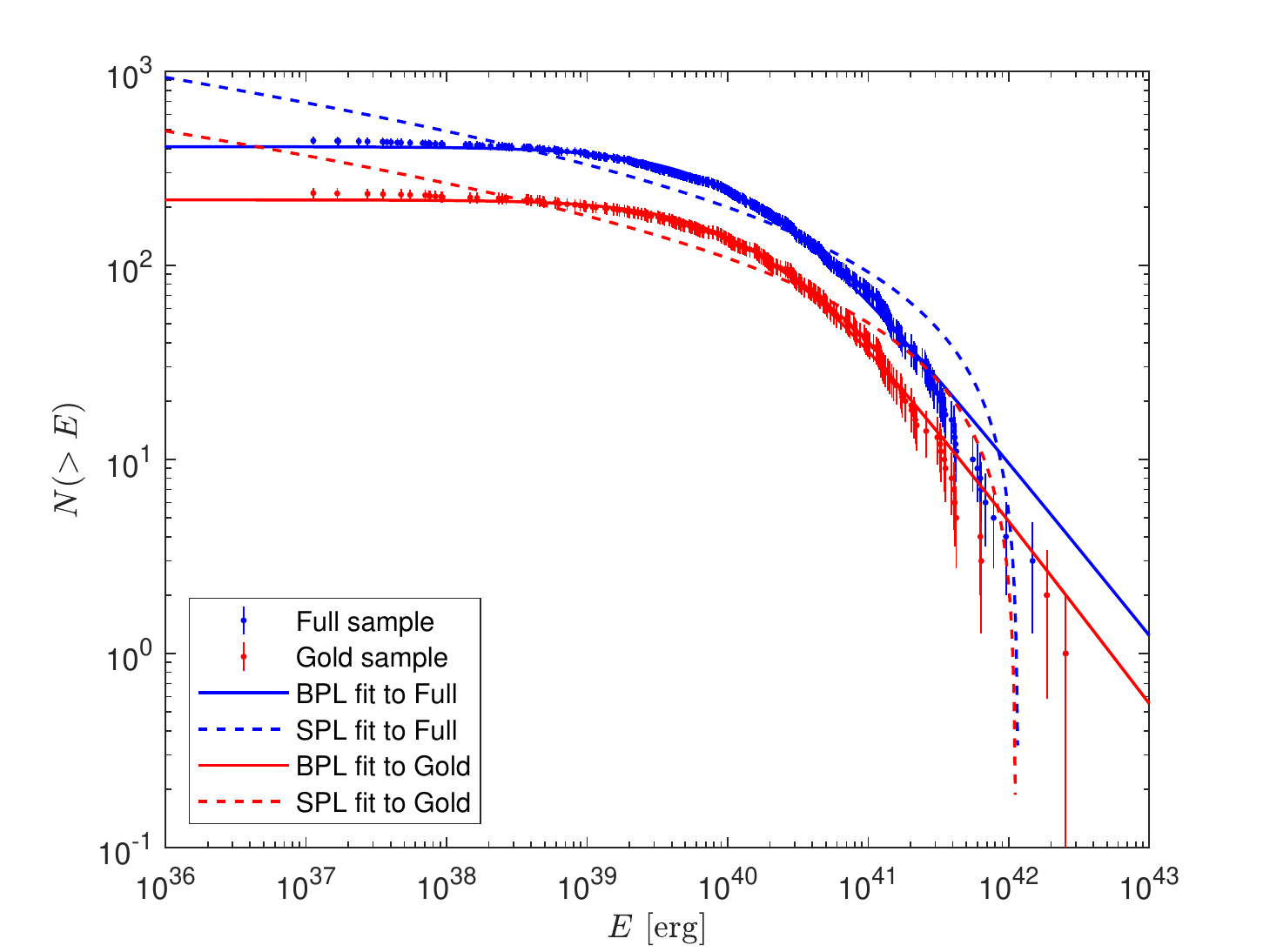}
 \caption{The cumulative distribution of fluence (left panel) and isotropic energy (right panel) of the non-repeating CHIME/FRBs with ${\rm DM_E}>100~{\rm pc~cm^{-3}}$. The solid and dashed lines are the best-fitting BPL model and SPL model, respectively.}\label{fig:cdf}
\end{figure}

Several works show that the distributions of fluence and energy of repeating FRBs follow simple power law (SPL) \cite{Wang:2016lhy,Wang:2019sio}. To check if the fluence and energy of the apparently non-repeating FRBs follow the same distribution or not, we calculate the cumulative distributions of fluence and energy of the non-repeating CHIME/FRBs (for both the Full sample and the Gold sample), and plot the results in Figure \ref{fig:cdf}. We try to fit the cumulative distributions of fluence and energy using the SPL model,
\begin{equation}
  {\rm SPL}:~~~N(>x)\propto(x^{-\beta}-x_c^{-\beta}),~~~x<x_c,
\end{equation}
where $x_c$ is the cut-off value above which the FRB count is zero. The uncertainty of $N$ is given by $\sigma_N=\sqrt{N}$ \cite{Wang:2019sio}. The best-fitting parameters are summarized in Table \ref{tab:cdf_parameter}, and the best-fitting lines are shown in Figure \ref{fig:cdf} (the dashed lines). As can be seen, for both the Full sample and the Gold sample, the SPL model fails to fit the distributions of fluence and energy. Especially at the left end, where the model prediction much exceeds the data points.

\end{multicols}
\begin{multicols}{2}

\end{multicols}
\begin{table}[htbp]
\centering
\caption{\small{The best-fitting parameters of the cumulative distributions of fluence and energy for the Full sample and the Gold sample.}}\label{tab:cdf_parameter}
{\begin{tabular}{lllll} 
\hline\hline 
\multirow{2}{*}{Fluence (Full)}
& SPL & $\beta=0.54\pm0.02$  & $x_c=66.30\pm3.52$ Jy ms & $\chi^2/{\rm dof}=7.48$ \\  
& BPL & $\gamma=1.55\pm0.01$ & $x_b=3.36\pm0.04$ Jy ms  & $\chi^2/{\rm dof}=0.23$ \\
\hline
\multirow{2}{*}{Fluence (Gold)}
& SPL & $\beta=0.48\pm0.03$  & $x_c=58.59\pm4.02$ Jy ms & $\chi^2/{\rm dof}=5.79$ \\  
& BPL & $\gamma=1.65\pm0.02$ & $x_b=3.96\pm0.07$ Jy ms  & $\chi^2/{\rm dof}=0.29$ \\
\hline
\multirow{2}{*}{Energy (Full)}
& SPL & $\beta=0.09\pm0.01$  & $x_c=(1.17\pm0.06)\times 10^{42}~{\rm erg}$ & $\chi^2/{\rm dof}=11.10$ \\  
& BPL & $\gamma=0.90\pm0.01$ & $x_b=(1.55\pm0.02)\times 10^{40}~{\rm erg}$  & $\chi^2/{\rm dof}=0.50$ \\
\hline
\multirow{2}{*}{Energy (Gold)}
& SPL & $\beta=0.08\pm0.01$  & $x_c=(1.13\pm0.09)\times 10^{42}~{\rm erg}$ & $\chi^2/{\rm dof}=7.12$ \\  
& BPL & $\gamma=0.95\pm0.01$ & $x_b=(1.82\pm0.04)\times 10^{40}~{\rm erg}$  & $\chi^2/{\rm dof}=0.29$ \\
\hline
\end{tabular}}
\end{table}
\begin{multicols}{2}

Lin $\&$ Sang \cite{Lin:2019ldn} showed that the bent power law (BPL) model fits the distributions of fluence and energy of the repeating burst FRB121102 much better than the SPL model. The BPL model takes the form
\begin{equation}
  {\rm BPL}:~~~N(>x)\propto\left[1+\left(\frac{x}{x_b}\right)^\gamma\right]^{-1},~~~x>0,
\end{equation}
where $x_b$ is the median value of $x$, i.e. $N(x>x_b)=N(x<x_b)$. The BPL model has a flat tail at $x\ll x_b$, and behaves like SPL model at $x\gg x_b$. The BPL model was initially used to fit the power density spectra of gamma-ray bursts \cite{Guidorzi:2016ddt}. Then it was shown that the BPL model can well fit the distribution of fluence and energy of soft-gamma repeaters \cite{Chang:2017bnb,Sang:2021cjq}. The choice of BPL model is inspired by the fact that the cumulative distributions of fluence and energy has a flat tail at the left end, as can be seen from Figure \ref{fig:cdf}. We therefore try to fit the cumulative distributions of fluence and energy of CHIME/FRBs using the BPL model. The best-fitting parameters are summarized in Table \ref{tab:cdf_parameter}, and the best-fitting lines are shown in Figure \ref{fig:cdf} (the solid lines). We see that the BPL model fits the data (both the Full sample and the Gold sample) much better than the SPL model. The BPL model fits the distribution of fluence very well in the full range. For the distribution of energy, the BPL model also well fits the data, except at the very high energy end.

\section{Discussion and conclusions}\label{sec:conclusions}

In this paper, we reconstructed the ${\rm DM_E}-z$ relation from 17 well-localized FRBs at $z<1$ using Bayesian inference method. The host DM was assumed to follow log-normal distribution with mean $\exp(\mu)$ and variance $\sigma_{\rm host}$, and the variance of DM of IGM was assumed to be redshift-dependent ($\sigma_{\rm IGM}=Fz^{-1/2}$). The free parameters were tightly constrained by 17 well-localized FRBs: $F=0.32_{-0.10}^{+0.11}$, $\exp(\mu)=102.02_{-31.06}^{+37.65}~{\rm pc~cm^{-3}}$ and $\sigma_{\rm host}=1.10_{-0.23}^{+0.31}$. These parameters are well consistent with that of Macquart et al. \cite{Macquart:2020lln}, who obtained $F=0.31_{-0.16}^{+0.13}$, $\exp(\mu)=68.2_{-35.0}^{+59.6}~{\rm pc~cm^{-3}}$ and $\sigma_{\rm host}=0.88_{-0.45}^{+0.65}$ from five well-localized FRBs. As the enlargement of FRB sample and one less free parameter ($\Omega_b$), our constraint is more stringent than that of Macquart et al. \cite{Macquart:2020lln}. We directly extrapolated these parameters to high redshift and reconstructed the ${\rm DM_E}-z$ relation up to $z=4$.

We further used the ${\rm DM_E}-z$ relation to infer the redshift of the first CHIME/FRB catalog. We found that the extragalactic DM of the non-repeating CHIME/FRBs follows CPL distribution, with a peak at $250$ pc cm$^{-3}$. The inferred redshift of the non-repeating CHIME/FRBs can also be fitted by CPL distribution, but with a significant excess at the low redshift range $0<z<0.1$, which may be caused by selection effect. We applied a set of criteria to exclude events which are susceptible to selection effect, as was described in Amiri et al. \cite{CHIMEFRB:2021srp}. We found that the extragalactic DM, as well as redshift of the remaining FRBs (which we call the Gold sample) still follow CPL distribution, and the excess at low redshift still exists. We further used the inferred redshift to calculate the isotropic energy of the non-repeating CHIME/FRBs. It was found that the distributions of energy and fluence can be well fitted by BPL model, with power index $\gamma=0.90\pm 0.01$ and $\gamma=1.55\pm 0.01$ for energy and fluence, respectively. However, the SPL model fails to fit both the distributions of fluence and energy, even for the Gold sample. The statistical properties of the non-repeating CHIME/FRBs are similar to that of the bursts from the repeating FRB source, FRB121102 \cite{Lin:2019ldn}. The BPL model has a flat tail at low-energy (low-fluence) end, thus it predicts much less dim bursts than the SPL model. The flatness at low-energy (low-fluence) end can be explained by the observational incompleteness, since some dim bursts may missing from detection. Note that the BPL model reduces to SPL model at high energy end, $N(>E)\propto E^{-\gamma}$. The power-law index of the energy accumulative distribution is $\gamma \approx 0.9$, which corresponding to $\hat\gamma\approx 1.90$ for the differential distribution. Interestingly, the power-law index of the non-repeating CHIME/FRBs is similar to that of repeating bursts from the single source FRB 121102, with $\hat\gamma\approx1.6\sim1.8$ \cite{Wang:2019sio}.

We emphasis that the CPL distribution of redshift is not intrinsic. The intrinsic redshift distribution should take into account the selection effect of the detector. Due to the lack of well-localized FRBs, the intrinsic redshift distribution is still poorly known. Several possibilities have been discussed in literature, such as the distribution similar with gamma-ray bursts \cite{Yu:2017beg}, a constant comoving number density with a Gaussian cutoff \cite{Li:2019klc}, the SFR history model \cite{Zhang:2020ass}, the modified SFR history model \cite{James:2021oep}, the compact star merger model with various time delay \cite{Zhang:2020ass}. In a recent work, Qiang et al. \cite{Qiang:2021ljr} considered several modified SFR history models and found that many of them are well consistent with the observed data of the first CHIME/FRB catalog, as long as the model parameters are chosen properly, but the simple SFR history model was fully ejected by the data. Hackstein et al. \cite{Hackstein:2020mxc} have investigate three different intrinsic redshift distribution models (constant comoving density model, SFR history model, and stellar mass density model). After considering the selection effects of CHIME telescope, them showed that the distribution of the observed redshift should have CPL shape. It remains to be a future work to study which model fits the CHIME/FRB best. In addition, Shin et al. \cite{Shin:2022crt} have studied the FRB population by assuming a Schechter luminosity function, and after calibrating the selection effects, they also found that the distribution of redshift has CPL shape.

When reconstructing the ${\rm DM_E}-z$ relation, it is important to reasonably deal with the ${\rm DM_{host}}$ term. The simplest way is to assume that ${\rm DM_{host}}$ is a constant \cite{Yu:2017beg,Wu:2020jmx,Qiang:2021ljr}. Of course, this is inappropriate because the actual value can vary significantly from bursts to bursts. Luo et al. \cite{Luo:2018tiy} parameterized ${\rm DM_{host}}$ as a function of SFR. However, statistical analysis of the well-localized FRBs shown that there is not strong correlation between ${\rm DM_{host}}$ and the host galaxy properties, including SFR \cite{Lin:2022afm}. Because there is a lack of direct observation on ${\rm DM_{host}}$, at present the most reasonable way is to model it using a probability distribution. Theoretical analysis and numerical simulations show that the probability of ${\rm DM_{host}}$ can be modeled by log-normal distribution with mean value $\mu$ and deviation $\sigma_{\rm host}$ \cite{Macquart:2020lln,Zhang:2020mgq}. Based on the IllustrisTNG simulation, Zhang et al. \cite{Zhang:2020mgq} showed that $\exp(\mu)$ has a power-law dependence on redshift, and the power-law index for repeating and non-repeating FRBs is slightly different. However, we found no evidence for the redshift evolution of $\exp(\mu)$ here. The median value of ${\rm DM_{host}}$ for the well localized FBRs we obtained here is about $\exp(\mu)\sim 100~{\rm pc~cm^{-3}}$. It is consistent with the ${\rm DM_{host}}$ of FRB20190608B ($\sim 137\pm 43~{\rm pc~cm^{-3}}$) obtained from optical/UV observations \cite{Chittidi:2021mfk}.

Due to the lack of high-redshift FRBs, the uncertainty on the ${\rm DM_E}-z$ relation is large at high redshift. The uncertainty mainly comes from the uncertainties on ${\rm DM_{IGM}}$ and ${\rm DM_{host}}$. The uncertainty on ${\rm DM_{IGM}}$ at redshift $z=1$ is about $\delta{\rm DM_{IGM}}\approx 0.3{\rm DM_{IGM}}\approx 270~{\rm pc~cm^{-3}}$. From the lognormal distribution, the uncertainty of ${\rm DM_{host}}$ is estimated to be $\delta {\rm DM_{host}}=\exp(\mu+\sigma_{\rm host}/2)(\exp(\sigma_{\rm host}^2)-1)^{1/2}\approx 200~{\rm pc~cm^{-3}}$, where $\exp(\mu)\approx 100~{\rm pc~cm^{-3}}$ and $\sigma_{\rm host}\approx 1$. The uncertainties of ${\rm DM_{MW}}$ and ${\rm DM_{halo}}$ is expected to be much smaller than that of ${\rm DM_{IGM}}$ and ${\rm DM_{host}}$, thus they have been ignored. We also ignored the DM of the FRB source, which is hard to model due the lack of knowledge on the local environment of FRBs. With the present knowledge we do not clearly know the probability distribution of ${\rm DM_{source}}$. In some models involving the merger of compact binary, this term is expected to be small \cite{Margalit:2019hke,Wang:2020aut}. Therefore, in most work this term is directly neglected. If ${\rm DM_{source}}$ does not strongly vary from burst to burst (such that it can be treated approximately as a constant), it can be absorbed into the ${\rm DM_{host}}$ term, while the probability distribution $p_{\rm host}$ does not change except for an overall shift. In this case, the parameter $\exp(\mu)$ should be explained as the median value of the sum of ${\rm DM_{host}}$ and ${\rm DM_{source}}$. Therefore, if ${\rm DM_{source}}$ does not vary significantly, including it or not should not affect our results. Another uncertainty comes from the parameter $f_{\rm IGM}$. In general case, $f_{\rm IGM}$ should be treated as a free parameter, together with $F$, $\exp(\mu)$ and $\sigma_{\rm host}$. But due to the small FRB sample, free $f_{\rm IGM}$ will lead to unreasonable result. So we fix $f_{\rm IGM}=0.84$ based on other independent observations. This will lead to the underestimation on the uncertainty of ${\rm DM_E}-z$ relation.

The conclusions of our paper are based on the assumption that the ${\rm DM}_E-z$ relation obtained from low-redshift data can be extrapolated to high redshift region. As is demonstrated in section 2, there is no strong evidence for the redshift dependence of host DM, at least at low-redshift region $z\lesssim 1$. But we can't prove this assumption at high redshift region, since there is lack of data points at $z>1$. So we just simply extrapolate the ${\rm DM}_E-z$ relation to high redshift region without proving it. Recent works \cite{James:2021jbo,James:2021oep} show that the ${\rm DM}_E-z$ relation may be nonmonotonic, with a turn point at a certain redshift. This is because that a FRB at low redshift is easier to be detected than at high redshift, for a given intrinsic luminosity. Therefore, a highly dispersed FRB is mainly caused by large DM of host galaxy, rather than by high redshift. For example, the large DM of FRB20190520B (${\rm DM_{obs}}\approx 1200~{\rm pc~cm}^{-3}$, $z\approx 0.241$) main attributes to the large value of ${\rm DM_{host}}$ ($\approx 900~{\rm pc~cm}^{-3}$) \cite{Niu:2021bnl}. Therefore, the uncertainty of ${\rm DM}_E-z$ relation we obtained in our paper may be significantly underestimated. We hope that the uncertainty can be reduced if more high-redshift FRBs are detected in the future.

\section*{Online material}
The parameters of the first (non-repeating) CHIME/FRB catalog are listed in a long table in the online material.

\end{multicols}

\vspace{-1mm}
\centerline{\rule{80mm}{0.5pt}}
\vspace{2mm}

\begin{multicols}{2}

\bibliographystyle{cpc-hepnp-0}
\bibliography{reference}

\end{multicols}

\section*{Online Material for ``Inferring redshift and energy distributions of fast radio bursts from the first CHIME/FRB catalog"}

The parameters of the first (non-repeating) CHIME/FRB catalog are listed in Table \ref{tab:chime_frbs} [The CHIME/FRB Collaboration, ApJS 257:59 (2021)]. Column 1: FRB name; Columns 2 and 3: the right ascension and declination of FRB source on the sky; Column 4: the observed DM; Column 5: the DM of the Milky Way ISM calculated using the NE2001 model; Column 6: the extragalactic DM calculated by subtracting ${\rm DM_{\rm MW}}$ and ${\rm DM_{\rm halo}}$ from the observed ${\rm DM_{\rm obs}}$, assuming ${\rm DM_{\rm halo}}=50~{\rm pc~cm^{-3}}$ for the Milky Way halo; Column 7: the observed fluence; Column 8: the inferred redshift; Column 9: the isotropic energy; Column 10: the flag for Gold sample (flag=1) or not (flag=0). The redshift and energy are only calculated for FRBs with ${\rm DM_E>100~pc~cm^{-3}}$. Note that the fluence and energy are the lower limits, since the fluence is measured assuming each FRB is detected at the location of maximum sensitivity.

\renewcommand\arraystretch{1.2} 
\begin{longtable}{cccccccccc}
\caption{The parameters of the first CHIME/FRB catalog.}\label{tab:chime_frbs}\\	
\hline  
FRBs & RA & Dec & ${\rm DM_{obs}}$ & ${\rm DM_{MW}}$ & ${\rm DM_E}$ & Fluence & $z_{\rm inf}$ & $\log(E/{\rm erg})$ & flag \\
& [$^{\circ}$ ] & [$^{\circ}$ ] & [${\rm pc/cm^{3}}$] & [${\rm pc/cm^{3}}$] & [${\rm pc/cm^{3}}$] & [Jy ms] & & & \\
\hline
\endfirsthead
\multicolumn{10}{c}{\textit{\---- Continued from previous page}} \\
\hline  
FRBs & RA & Dec & ${\rm DM_{obs}}$ & ${\rm DM_{MW}}$ & ${\rm DM_E}$ & Fluence & $z_{\rm inf}$ & $\log(E/{\rm erg})$ & flag\\
& [$^{\circ}$ ] & [$^{\circ}$ ] & [${\rm pc/cm^{3}}$] & [${\rm pc/cm^{3}}$] & [${\rm pc/cm^{3}}$] & [Jy ms] & & & \\
\hline
\endhead
\hline
\endfoot
\hline
\endlastfoot

20180725A & $93.42$ & $67.07$ & $716.6$ & $72.4$ & $594.2$ & $4.10\pm2.30$ & $0.668_{-0.315}^{+0.163}$ & $40.418_{-0.973}^{+0.404}$ & $0$\\
20180727A & $197.72$ & $26.42$ & $642.1$ & $21.2$ & $570.9$ & $2.31\pm0.76$ & $0.641_{-0.340}^{+0.159}$ & $40.129_{-0.900}^{+0.338}$ & $0$\\
20180729A & $199.40$ & $55.58$ & $108.4$ & $29.6$ & $28.8$ & $17.00\pm10.00$ & $\--$ & $\--$ & $0$\\
20180729B & $89.93$ & $56.50$ & $318.6$ & $95.4$ & $173.2$ & $1.20\pm0.74$ & $0.104_{-0.081}^{+0.118}$ & $38.136_{-1.737}^{+0.906}$ & $0$\\
20180730A & $57.39$ & $87.19$ & $849.2$ & $59.5$ & $739.7$ & $27.00\pm12.00$ & $0.848_{-0.332}^{+0.183}$ & $41.468_{-0.737}^{+0.348}$ & $0$\\
20180801A & $322.53$ & $72.72$ & $656.7$ & $91.1$ & $515.6$ & $7.90\pm4.80$ & $0.570_{-0.313}^{+0.149}$ & $40.549_{-1.171}^{+0.431}$ & $0$\\
20180806A & $228.56$ & $75.62$ & $739.2$ & $39.9$ & $649.3$ & $7.60\pm6.60$ & $0.736_{-0.306}^{+0.167}$ & $40.780_{-1.401}^{+0.469}$ & $0$\\
20180810A & $101.47$ & $34.86$ & $415.7$ & $105.5$ & $260.2$ & $1.70\pm1.00$ & $0.224_{-0.166}^{+0.132}$ & $38.990_{-1.617}^{+0.640}$ & $0$\\
\end{longtable}

\end{document}